# Ferroelastic twin wall mediated ferro-flexoelectricity and bulk photovoltaic effect in SrTiO₃

Ri He[1#], Haowei Xu[2#], Peijun Yang[1], Kai Chang[3], Hua Wang[2, 3, 4*], Zhicheng Zhong[1,5*]

[1]Key Laboratory of Magnetic Materials Devices & Zhejiang Province Key Laboratory of Magnetic Materials and Application Technology, Ningbo Institute of Materials Technology and Engineering, Chinese Academy of Sciences, Ningbo 315201, China

[2]Department of Nuclear Science and Engineering, Massachusetts Institute of Technology, Cambridge, MA, USA

[3]School of Physics, Zhejiang University, Hangzhou 310058, China

[4]ZJU-Hangzhou Global Scientific and Technological Innovation Center, Hangzhou 311215, China

[5]China Center of Materials Science and Optoelectronics Engineering, University of Chinese Academy of Sciences, Beijing 100049, China

## Abstract

Ferroelastic twin walls in nonpolar materials can give rise to a spontaneous polarization due to symmetry breaking. Nevertheless, the bi-stable polarity of twin walls and its reversal have not yet been demonstrated. Here, we report that the polarity of SrTiO₃ twin walls can be switched by ultra-low strain gradient. Using first-principles-based machine-learning potential, we demonstrate that the twin walls can be deterministically rotated and realigned in specific directions under strain gradient, which breaks the inversion symmetry of a sequence of walls and leads to a macroscopic polarization. The system can maintain polarity even after the strain gradient is removed. As a result, the polarization of twin walls can exhibit ferroelectric-like hysteresis loop upon cyclic bending, namely ferro-flexoelectricity. Finally, we propose a scheme to experimentally detect the polarity of twin wall by measuring the bulk photovoltaic responses. Our findings suggest a twin-wall-mediated ferro-flexoelectricity in SrTiO₃, which could be potentially exploited as functional elements in nano-electronic devices design.

---

# These authors contribute equally to this work
* daodaohw@zju.edu.cn
* zhong@nimte.ac.cn

Domain walls are intriguing topological defects commonly found in ferroic materials, possessing unique structural and properties that markedly depart from those of bulk systems [1,2]. Among these, ferroelastic twin walls (TWs) stand out as highly promising candidates for emerging functionalities in nano-electronic devices, as they represent a two-dimensional nanosized feature within a homogeneous solid [3]. Unlike conventional domain walls, TWs only occur in two tetragonal domains whose elongate axes are mutually perpendicular. This breaks local inversion symmetry, leading to the natural emergence of spontaneous polarization, even in a nonpolar bulk material [4-6]. An illustrative example of this phenomenon can be found in $SrTiO_3$, an essential material for oxide electronics. The polarity of TWs in $SrTiO_3$ has been experimentally confirmed through the observation of electromechanical resonance peak [7-9]. However, direct experimental observation of TW polarity in $SrTiO_3$ remains challenging due to the opposite of the polarization in neighboring TWs. Theoretically, the Ginzburg-Landau phenomenological and effective Hamiltonian models suggest that such polarity at TWs originates from the strain gradient [10,11], oxygen octahedron tilts gradient [12], and Ti atoms displacement [13]. The strain gradient origin has also been confirmed by atomistic simulations based on a simplified toy model [14].

Although ferroelastic TWs in $SrTiO_3$ have been extensively studied experimentally and theoretically over the last decade, several fundamental issues about their properties remain unsettled [1]. One of the most significant issues is the determination of whether the TWs exhibit bi-stable polarity states, and if so, how this polarity can be detected, manipulated, and switched by external fields. Another critical aspect to address is the control TWs' dynamics.

To better tackle above two issues, it has become imperative to employ atomic simulations based on first-principles density functional theory (DFT). However,conducting DFT simulations of periodic sequences of parallel TWs are challenging due to their high computational cost, which hinders the thorough investigation of the coupling between structural order parameters (e.g., polarization, strain, and oxygen octahedron tilt) near the periodic neighboring walls. In addition, the breakdown of crystal periodicity in an inhomogeneous domain wall system is challenging for periodic DFT calculation. While classical molecular dynamics offers computational efficiency, it falls short in accurately capturing the subtle energy variation induced by antiferrodistortion in tetragonal $SrTiO_3$. Fortunately, recent advances in machine-learning-based potential have opened avenues for overcoming these limitations[15-19]. In recent works, we have developed a machine learning based deep potential (DP) model for $SrTiO_3$ using training dataset obtained from DFT calculation [16,17]. Our DP model demonstrates excellent agreement with DFT results across a wide range of bulk properties, including antiferrodistortive structure, total energies, atomic forces, elastic constants, and phonon dispersion relation.

Herein, we further improve our DP model by incorporating additional freestanding configurations with varying thicknesses into our existing training dataset, so that it can describe various complex configurations in $SrTiO_3$, such as surface with $TiO_2$ and $SrO$ termination. Subsequently, we perform the large-scale atomic simulations of $SrTiO_3$ TWs based on deep potential molecular dynamics (DPMD) simulations. We find that the TWs corrugate the surface on (0 0 1) membrane because of lattice misfit. The polarity of TW is always parallel to the TWs and has positive (negative) value in convex (concave) surface, resulting in the opposite polarization directions in neighboring TWs and a resultant macroscopic polarization that effectively vanishes. More importantly, we reveal the polarity of TW exhibits bi-stable states, which can be switched by strain-gradient-induced rotation of the TWs. This wall rotation introduces a breakdown of macroscopic inversion symmetry in a periodic sequence of TWs, resulting a macroscopic polarization. Remarkably, this system can maintain polarity even when strain gradient is removed. As a result, the polarization of TWs can exhibit ferroelectric-like hysteresis loop during cyclic bending，a phenomenon termed ferro-flexoelectricity. Our findings suggest an extrinsic TW mediated ferro-flexoelectricity in $SrTiO_3$, which could be potentially exploited as functional elements in material design and applications.

***Twin wall structure and polarization properties***. At low temperatures, $SiTiO_3$ has a tetragonal phase with *I4/mcm* space group. Ferroelastic TWs occur in two tetragonal domains whose elongate axes are perpendicular to each other. We adopt a twenty-atom tetragonal unit cell to construct a $1 \times 1 \times 20$ supercell and then create the two TWs using a twin-like operation (Supplementary Material S3 and Fig. S1 [20]). The fully DFT relaxed supercell with TWs is shown in Fig. S2a (see Method in Supplementary Material [20]). In consideration of the delicate of lattice distortion in TW region, we plot an exaggerated schematic of the supercell with two antisymmetric TWs parallel to (1 0 1) planes in Fig 1a. Interestingly, the supercell shows a periodic structural corrugation, the concave (180.58°) and convex (179.42°) twin angles are formed at two antisymmetric TW regions in (-1 0 1) plane (see Fig 1a). These twin angles also cause corrugations in the (0 0 1) surface of membrane (see Fig. S3d [21]). Such spontaneous surface corrugation originates from lattice matching of the different extension of their tetragonal unit cells at TWs, as sketched in Fig.2b. Such (0 0 1) surface corrugation can be detected by optical reflection microscopy as a striped contrast experimentally [22] and then its angle has been indirectly measured by Honig et al [23]. The wall energy of TW is much lower than that of antiphase hard and easy wall (see Supporting Information S3 and Fig. S6 [20]), demonstrating that TW is the most stable wall configuration in $SrTiO_3$.

The tilt pattern of tetragonal $SrTiO_3$ is $a^0a^0c^-$ in Glazer notation[24], and the atomic structural properties of TW are characterized by structural order parameter of axial vector $\varphi$ from the staggered rotations of $TiO_6$ octahedra as sketched by inset of Fig. 1b. The vector $\varphi$ is projected along two directions, either perpendicular ($\varphi_z$) or parallel ($\varphi_x$) to the wall. Figure 1b shows the $\varphi$ profile across the TW, the head to head (HH) TW can be viewed as a transition area where $\varphi_x$ passes from $\varphi_x$ to $-\varphi_x$, while $\varphi_z$ changes from $\varphi_z$ to $-\varphi_z$ across the head to tail (HT) wall (Fig. S2a and b [20]). This indicates that when moving across the walls, one of the two components of $\varphi$ should be zero. According to Landau-type free-energy model[10,11], the polarization ($P$) does not occur in the bulk region because of the repulsive interaction between $P$ and $\varphi$. Nevertheless, the vanishing of one of the $\varphi$ components implies a possible polarization $P$ inside the wall.

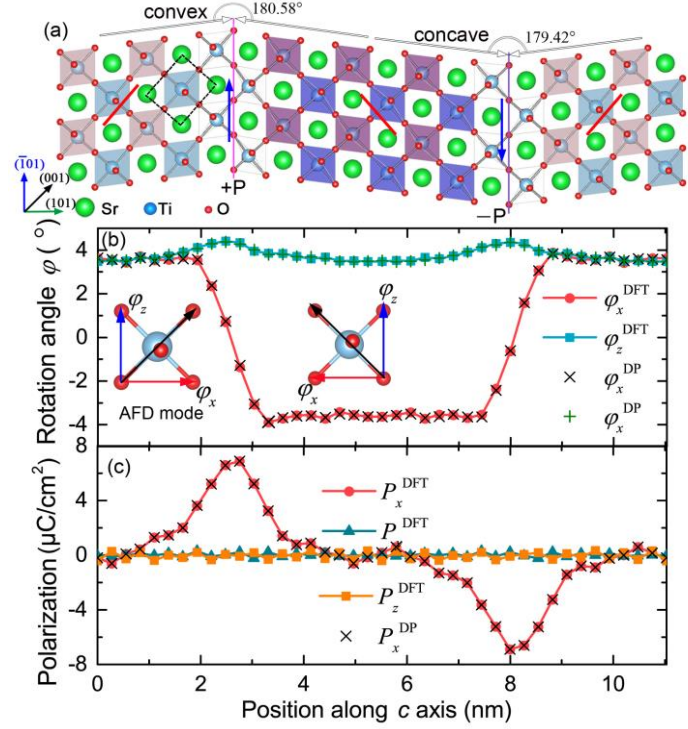

Figure 1: Twin wall structure and polarization profile in tetragonal phase SrTiO₃. (a) Schematics of atomistic structure of the twin wall (the tilting octahedral distortions is exaggerated). Blue arrows in twin wall indicate that the direction of spontaneous polarization is parallel to the [-101] direction. black dashed square indicates the unit cell used for calculation of polarization via Born effective charge. (b) Vector φ (staggered rotations of octahedra) and (c) polarization profile depending on the distance across the two antisymmetric twin walls.

Based the analysis on structural order parameter $\varphi$ and polarization ($P$) profiles (for details, see supplementary material S4), the polar TW in SrTiO₃ has three features: (*i*) The TW has had a wide width of ~8.2 nm (Fig. S4 [20]), in excellent agreement with the value of 7~8 nm predicted by atomistic Hamiltonian model[13]; (*ii*) $P$ is pronounced along the *x* axis, reaching a value of ±7.1 μC/cm² at the centre of TWs. The other two components of $P_y$ and $P_z$ are negligible (Fig. 1c and Fig. S2d [20]). Nevertheless, $P$ changes its sign from one wall to the other, forming antiferroelectric-like configuration along *x* axis (Fig. 1a and Fig. S3a [20]). It indicates that the paired antisymmetric TWs within the supercell maintain charge neutrality, thus the macroscopic $P$ would vanish with a sequence of parallel TWs system; (*iii*) The polarization is always parallel to the TW and its orientation is related to the corrugation in (-1 0 1) plane (i.e., the orientation of elongate axes respect to the walls). Specifically, $P$ has positive (negative) value in convex (concave), as marked by blue arrows in Fig. 1a.

***Twin-wall-rotation-induced switchable polarization.*** TWs can give rise to a spontaneous polarization, yet whether the polarization possesses bi-stable states, and the methods for manipulating, detecting, and reorienting the polarization remain undisclosed. To address this intriguing question, we used a simplified point charge lattice model to reveal the fundamental lattice-derived origins of polarity, as illustrated Figs. 2a, b [20]. The inversion symmetry is broken by the lattice matching, aligning with the distinct orientations of tetragonal domains at walls. This alignment triggers relative displacements of the centers of negative and positive charges, resulting in polarization. Once the orientations of tetragonal domains with respect to wall are fixed, the spontaneous polarization $P$ cannot be switched by electric field. However, Fig. 2c shows that if the orientation of TW rotates 90° with respect to the elongate axes of adjacent tetragonal domains, the polarity can be effectively switched. This indicates that the polarity in TW has potential reversible bi-stable states, which can be manipulated by adjusting the orientation of the wall with respect to the elongate axes.

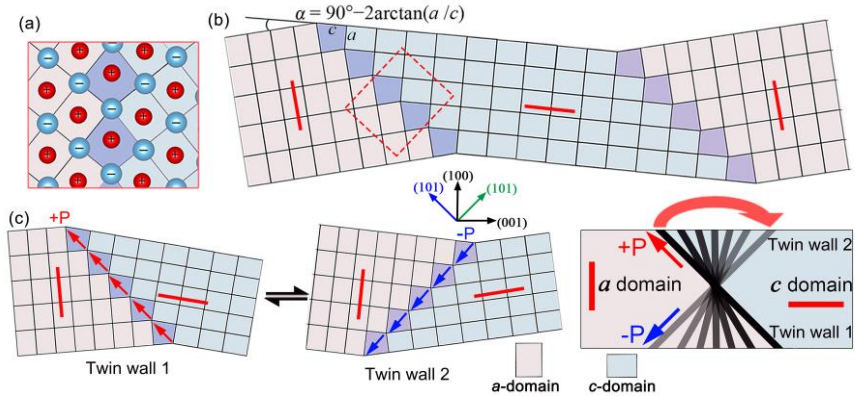

Figure 2: (a) Schematic representation of the domain and twin wall structure in a film with *a-c* domains and corresponding simplified point charge lattice model from red dashed square in (b). (b) The surface is angled at the intersection between a (blue) and c (red) domains with an angle of tan(α) = (*c/a*). Lattice model indicates that the lattice misfit of the different orientation of tetragonal domains at walls breaks inversion symmetry. The red lines represent the orientation of elongate axes. (c) If the orientation of the wall rotates by 90° with respect to the elongate axes of adjacent tetragonal domains, then the polarization $P$ can be switched. This indicates that the polarity in the TW can exist in reversible bi-stable states, which can be turned by wall's orientation.

The next challenge is to determine how external fields can be used to manipulate the orientation and polarity of the wall. Previous experimental work has shown that the TW in pure STO exhibit virtually no pinning and maintain high mobility under [1 0 0] uniaxial strain below 105K, because TW does not contain any vacancy or dislocation[25]. To unveil the intricate relationship between the strain, polarity, and electromechanical properties, we simulated the evolution of axial vector $\varphi$ configurations in (0 0 1) quasi-2D superlattice with $106 \times 31 \times 4$ pseudo cubic units by DP model. It' noteworthy that the DP model can describe the surface of SrTiO$_3$ with DFT accuracy (see Supporting Information [20]). The DP optimized thin-film superlattice is shown in Fig. 3a. Here, we identified the lattices as $c$-domain with $\varphi$ along the $y$-axis (blue arrows) and $a$-domain with $\varphi$ along the $x$-axis (red arrows). The typical TW texture is shown by $\varphi$ map in Fig. 3b, which consists of stripes pattern aligned along the [-1 0 1] axes. The distribution of electric dipole in relaxed SrTiO$_3$ membrane superlattice is shown in Fig. S3a [20]. The opposite polarization (red and blue arrows) in neighboring parallel TWs cancel each other out, thus the macroscopic polarization vanishes in membrane. Furthermore, the high-accuracy atomic relaxation allows us to visually discern the delicate convex and concave corners in (0 0 1) surface (Fig. S3d [20]).

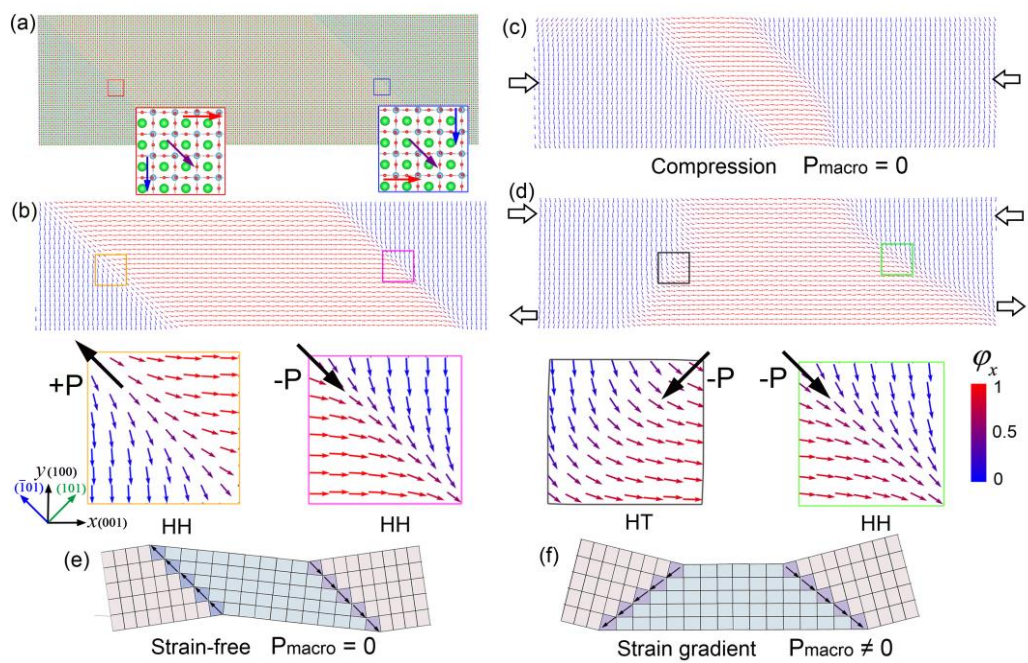

Fig. 3 The twin wall patterns of the SrTiO$_3$ membrane upon uniaxial compression and bending at 10 K. (a) The DP optimized 132604-atom supercell with 40.0 nm (length, vacuum) $\times$ 12.1 nm (thickness, vacuum) $\times$ 3 nm (period) containing two anti-symmetrical TWs. The insets show the local zoom-in atomic structure of two TWs. (b-d) The axial vector of staggered rotations of TiO6 octahedra ($\varphi$) configurations at strain-free, uniaxial compression, and bending strain. The blue and red arrows represent the vector $\varphi$ with components [1 0 0] and [0 0 1], respectively. (e-f) Schematic representation of geometry in membrane at strain-free and bending strain conditions.

The DPMD simulations show that uniaxial compressive strain can promote the domain switching, i.e. $a$-domain expands and $c$-domain shrinks, resulting in horizontal motions of wall (Fig. 3c). While the uniaxial tensile strain drive the wall to move in the opposite directions, consistent with previous phenomenological predictions [25]. However, it is important to note that these horizontal movements of TWs do not bring about a change in their polarity. Considering that the tensile and compressive strains lead to opposing horizontal movements of TWs, here, we propose that strain gradients may have the capability to rotate the TWs. To confirm this strain-dependent motion of TWs, we examine the evolution of atomic structure in supercell under bending at 10 K (see Methods in Supporting Information [20]). As we applied a slight strain gradient with 57.1 m$^{-1}$, the top half of supercell under compression led to expansion of $c$ domain, while bottom half under tension led to shrink of $c$ domain. Consequently, the left TW rotates 90˚ clockwise and the right TW is immobile as illustrated in Fig. 3d. TW rotation accompanied by its polarization switching, resulting a macroscopic polarization in the membrane.

To assess the shape and polarity maintaining characteristics of membranes under cyclic loading, we carried out the DPMD simulation of loading and unloading due to bending. Interestingly, the TW pattern cannot switch back upon strain gradient unloading (Fig. S5 [20]), demonstrating its stability. Previous work also reported that the BaTiO$_3$ membrane cannot achieve a complete shape recovery due to the irreversible local transformation of domain patterns[26]. Moreover, the residual bending strains are relaxed by the two convex (concave) corners on the top (bottom) surface (see Fig. S5 [20]). Therefore, the pure SrTiO$_3$ membrane with periodic TWs exhibits remarkable polarization-strain-gradient hysteresis loop as shown in Fig. 4a, and the insets show TW patterns in corresponding regions. In addition, it's worth noting that the two symmetric polarized states 1 and 2 have higher energy of ~1.7 meV/atom than that of nonpolar state 0 (Fig. 4b). Thus, it provides us with the possibility of "engineering" a macroscopic ferrielectric polarization in a periodic TW structure using strain gradient. To gain a deeper insight on polarity in configuration of Fig. 3(e) with strain gradient, we calculate the macroscopic $P$ and find it to be $4.91 \times 10^{-6}$ C/m² in [001] direction, when the strain gradient is 57.1 m⁻¹. Therefore, the flexoelectric coefficient $\mu_{12}$ is estimated to be $8.6 \times 10^{-8}$ C/m (see Method [20]), one order of magnitude larger than the experimental value $6 \times 10^{-9}$ C/m [27]. This discrepancy can be traced back to the differences in the density and distribution of walls between simulation and experiment. In experiment, the orientation in SrTiO₃ sample is random [9,22,28].

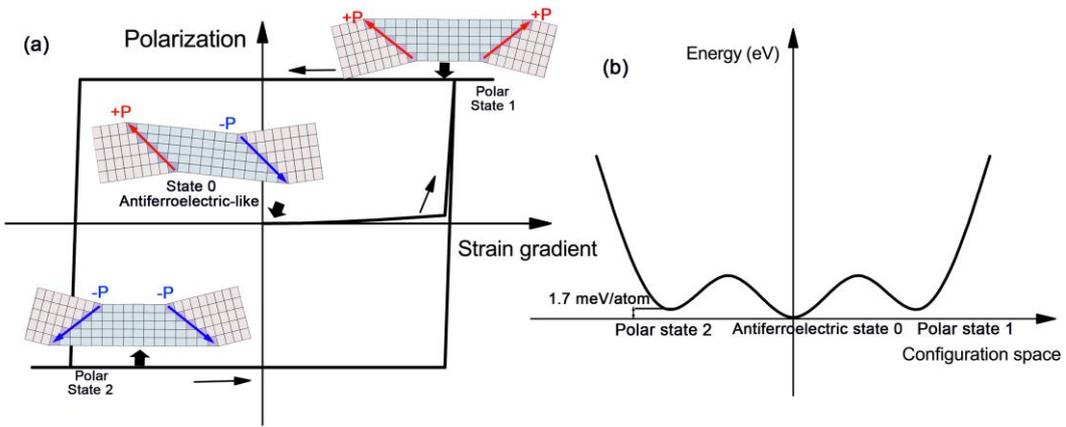

Fig. 4 (a) The pure SrTiO₃ thin-film with periodic TWs exhibits polarization-strain gradient hysteresis loop. The insets show the polar and non-polar states in corresponding regions. (b) The total energy profiles of transition from ono-polar to polar states.

***Bulk photovoltaic effect in ferroelastic twin wall.*** Detecting the polarization and reversal of TWs in SrTiO₃ by experiment is a tough task [1]. It is well known that nonlinear optical (NLO) responses are strongly dependent on the structural properties of the materials [29]. Particularly, second-order NLO effect, such as the bulk photovoltaic (BPV) effect are forbidden in centrosymmetric systems. Meanwhile, in ferroelectric materials, the direction of the BPV current can be controlled by switching the spontaneous electric polarization [30]. Interestingly, the TWs in SrTiO₃ have net electric polarizations as well, which break the original symmetry in SrTiO₃, leading to non-zero BPV responses. To demonstrate this effect, we build a supercell with two domain boundaries with opposite polarization along $x$ direction, and the local electric polarization density is shown in Fig. 1(c), where two spatial regions with opposite polarizations (labelled as $L = \uparrow$ and $\downarrow$, respectively) can be observed. Then, we calculate the local BPV conductivity $\sigma_{bc}^{a,L}(\omega)$ using[21,31],

$$\sigma_{bc}^{a}(\omega) = -\frac{e^2}{\hbar^2\omega^2} \int \frac{d\boldsymbol{k}}{(2\pi)^3} \sum_{mnl} \frac{f_{lm}v_{lm}^b}{\omega_{ml} - \omega + i/\tau} \left( \frac{j_{mn}^{a,L}v_{nl}^c}{\omega_{mn} + i/\tau} - \frac{v_{mn}^c j_{nl}^{a,L}}{\omega_{nl} + i/\tau} \right)$$

here $a$ $(b,c)$ indicates the direction of the photocurrent (the polarization of the light field). $L = \uparrow, \downarrow$ labels the spatial region. $f_{lm} \equiv f_l - f_m$ and $\omega_{lm} \equiv \omega_l - \omega_m$ are the occupation and energy difference between two electronic bands $l$ and $m$. $v_{nl} \equiv \langle n|\hat{v}|l\rangle$ is the velocity operator. The local current operator is defined as $j^{a,L} = P_L^\dagger j^a P_L$ with $j^a = -ev^a$. The projection operator is $P_L \equiv \sum_{i \in L} |\psi_i\rangle\langle\psi_i|$, where $|\psi_i\rangle$ is atomic orbitals located in the spatial region $L$. The explicit $\boldsymbol{k}$-dependence of the quantities are omitted. The carrier lifetime $\tau$ is set as 0.2 ps uniformly hereafter.

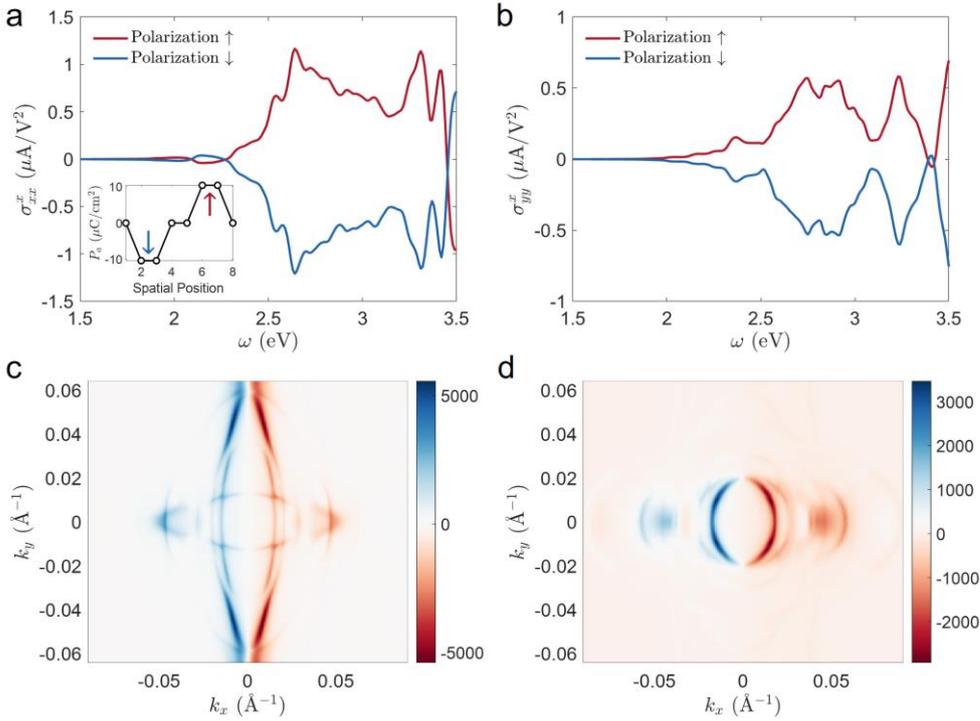

Fig. 5 Nonlinear optical (NLO) responses on twin wall: (a-b) Different components of the local BPV conductivity tensor as a function of light frequency ω. Inset of (a) shows the local electric polarization along x direction. Two regions with opposite polarizations are labelled with L=↑ and ↓. (c-d) show the contributions from each $\boldsymbol{k}$-point in the Brillouin zone to the local BPV conductivity in the ↑ polarization region. (c) and (d) are for $\sigma_{xx}^x$ and $\sigma_{yy}^x$, respectively. The colorbar is in arbitrary unit.

Two non-zero components of the $\sigma$ tensor for $L = \uparrow$ and $\downarrow$ are shown in Fig. 5a, b. One can see that $\sigma_{xx}^x$ and $\sigma_{yy}^x$, which are sensitive to the polarizations along $x$ direction, are nearly opposite for $L = \uparrow$ and $\downarrow$. This indicates that when linearly polarized light is applied, there will be currents travelling along the opposite directions near the domain boundaries. The contribution from each $k$-point in the Brillouin zone is shown in Fig. 5c, d. These results suggest that the BPV effect can serve as noninvasively method to detect the local polarization and its reversal in SrTiO$_3$ TWs. Notably, the BPV current is generated by the local electric polarization of light and does not necessitate any physical contact with the sample. As a result, it emerges as a promising technique for the nondestructive characterization of TWs in SrTiO$_3$. More importantly, the BPV effect can be harnessed to facilitate the flexo-photovoltaic effect induced by strain gradients. This opens up exciting possibilities for the use of TWs in applications such as energy harvesting and sensing.

To conclude, using machine-earning potential method, we demonstrate that a polarization is indeed confined to TWs in non-polar SrTiO$_3$, and that the TWs corrugate the (001) surface on membrane sample due to lattice mismatch. The polar state of wall is intricately linked to their geometrical corrugation. We confirm that the polarity of TWs exhibits bi-stable states and can be effectively switched through wall rotation driven by strain gradients. Based on the above results, we propose a mechanism for generating flexoelectricity by TW rotation. Specifically, a 90° rotation of walls results in a reversal of their polarization, leading to a breakdown of macroscopic inversion symmetry in a periodic sequence of walls and consequently inducing a macroscopic polarization. The calculated flexoelectric coefficient is one order of magnitude larger than the experimental value, demonstrating the contribution of such mechanism should not be ignored in low-temperature measurement. Our findings provide an essential understanding of spontaneous polarity within ferroelastic TWs and pave the way for their potential utilization as functional elements in next-generation nanoelectronics devices.

All the input files, final training datasets, and DP model files to reproduce the results contained in this paper are available in website (https://www.aissquare.com/).

This work was supported by the National Key R&D Program of China (Grants No. 2021YFA0718900 and No. 2022YFA1403000), the Key Research Program of Frontier Sciences of CAS (Grant No. ZDBS-LY-SLH008), the National Nature Science Foundation of China (Grants No. 11974365, No. 12204496, and No. 12304049), the K.C. Wong Education Foundation (GJTD-2020-11), and the Zhejiang Provincial Natural Science Foundation (LDT23F04014F01).

# Supporting Information

# Ferroelastic twin wall mediated ferro-flexoelectricity and bulk photovoltaic effect in SrTiO₃

Ri He[1#], Haowei Xu[2#], Peijun Yang[1], Kai Chang[3], Hua Wang[2,3,4*], Zhicheng Zhong[1,5*]

[1]Key Laboratory of Magnetic Materials Devices & Zhejiang Province Key Laboratory of Magnetic Materials and Application Technology, Ningbo Institute of Materials Technology and Engineering, Chinese Academy of Sciences, Ningbo 315201, China

[2]Department of Nuclear Science and Engineering, Massachusetts Institute of Technology, Cambridge, MA, USA

[3]School of Physics, Zhejiang University, Hangzhou 310058, China

[4]ZJU-Hangzhou Global Scientific and Technological Innovation Center, Hangzhou 311215, China

[5]China Center of Materials Science and Optoelectronics Engineering, University of Chinese Academy of Sciences, Beijing 100049, China

## S1. **Methods**

<u>Machine-learning potential of SrTiO$_3$</u>

Machine-learning is a powerful method for provide high-accuracy interaction potential[1]. The basic idea of machine-learning potential is construction of deep neural network and to fit the DFT calculation data of abundant configurations. For a well-trained DP model, given any large-scale configuration, it can figure out the corresponding total energy and atomic forces at DFT-level accuracy. Recently, we developed a machine-learning-based deep potential (DP) for SrTiO$_3$ using training dataset from DFT calculation[2,3]. A large number of representative configurations in training dataset were obtained by performing concurrent learning procedure[4]. The Our DP model has been proved to be in satisfactory agreement with DFT results in a wide range of bulk properties. However, our focus is on the studying the electromechanical property of twin wall at strain conditions, the twin walls break the crystal periodicity of supercell in [0 1 0] and [0 0 1] orientations as illustrated in Fig. 3. Therefore, the freestanding supercell with surface is needed, thus the training dataset requires inclusion of characteristic configurations of TiO$_2$ and SrO terminated surfaces. In this work, the freestanding SrTiO$_3$ films with thickness varying from 2 to 10 unit cells were introduced to concurrent learning procedure. The vacuum region spans 30 Å to prevent coupling between periodic images.

The concurrent learning procedure contains a series of iterations, and each iteration includes three steps: (1) training the DP model from initial training dataset, (2) exploring configurations by running NPT (deep potential molecular dynamics) DPMD simulations at different temperatures and pressure, and (3) labeling configurations and adding them to the training dataset, then repeating step (1) again. In the exploration step, DP model is used for MD simulations at various temperatures and biaxial strain to extend the configuration space. We start with $2 \times 2 \times n$ ($n = 2$ to 10) supercells of DFT-optimized freestanding (with TiO$_2$ and SrO terminal surface). After iterating this procedure 13 times, 2313 freestanding configurations were generated. The DP model is trained using these configurations and corresponding PBE-based DFT energies, with fitting deep neural network of size (240, 240, 240). The cutoff radius of the model is set to 6.5 Å for neighbor searching, while the smoothing function starts from 2.5 Å. The DEEPMD-KIT code is used for training of DP model[5]. The DP compression scheme was applied in this work for accelerating the computational efficiency of the DPMD simulations[6].

## **DFT calculations**

The initial training dataset is obtained by performing a 10-step *ab initio* MD simulation for randomly perturbed structures at 50 K. After labeling candidate configurations, self-consistent DFT should be performed. All DFT calculations were performed using a plane-wave basis set with a cutoff energy of 500 eV as implemented in the Vienna *Ab initio* Simulation Package (VASP)[7,8], and the electron exchange-correlation potential was described using the generalized gradient approximation and PBEsol scheme[9]. The Brillouin zone was sampled with a $6 \times 6 \times 1$ k-point grid for the $2 \times 2 \times n$ freestanding supercells.

We calculate the domain wall energies for several domain wall structures. The domain wall energy is defined as follows:

$$E_{wall} = \frac{E_{DWcell} - E_{bulk}}{2S},$$

where $E_{DWcell}$ represents the total energy of supercell with domain wall, $E_{bulk}$ is the energy of corresponding bulk system, $S$ is the area of wall.

Then a tight-binding (TB) Hamiltonian is constructed based on the plane-wave DFT results using the Wannier90 package[10]. The TB Hamiltonian is utilized to calculate the bulk photovoltaic conductivities. The BZ integration is carried out by **$k$**-mesh sampling with $\int \frac{dk}{(2\pi)^3} = \frac{1}{V} \sum_k w_k$, where $V$ is the volume of the supercell cell and $w_k$ is a weight factor.

## **Molecular dynamics simulations**

The MD simulations were carried out using LAMMPS code with periodic boundary conditions[11]. The atomistic interactions of SrTiO₃ are described by the developed machine-learning potential. The Nose-Hoover thermostat is employed to control temperature[12]. The velocity Verlet algorithm is used for integrating the equations of motion with a time step of 1.0 fs in all MD simulations.

To perform the large-scale atomic simulations on (0 0 1) membrane. The multidomain SrTiO₃ membrane superlattice with a tetragonal structure was constructed with the orientations of x-[1 0 0], y-[0 1 0] and z-[0 0 1], which contain a-domain, c-domain, and two twin walls. We construct a quasi-2D superlattice with $106 \times 31 \times 4$ pseudo cubic units. The vacuum region was used in x-[1 0 0] and z-[0 0 1] direction and periodicity was used in y-[0 1 0] direction. The DP optimized membrane superlattice is shown in Fig. S3. The convex and concave corners are found in (0 0 1) surface. To perform bending deformation, several atomic layers at both ends of the membrane were fixed rigidly as the loading grip. The bending was achieved by tilting the rigid loading ends against each other. The tilt was performed followed by atomic relaxation at the 10 K for 10 ps. The dipole moment was evaluated by averaging over the last 100 ps. The local pseudocubic cell polarization (**P**) are calculated by atomic displacements (ui) with respect to the referenced cubic phase multiplied by the Born effective charges ($Z_i^*$):

$$P = \sum Z_i^* u_i \,,$$

the components of Born effective charge tensors along the out-of-plane direction were obtained by the DFT calculations: $Z^*_{Sr} = 2.54$, $Z^*_{Ti} = 7.12$, $Z^*_{O1} = -5.66$, $Z^*_{O2} = -2.00$, where O1 denotes the oxygen atom in the SrO layer, and O2 denotes the oxygen atom in the TiO₂ layer. The distribution of electric dipole in relaxed SrTiO₃ membrane superlattice is shown in Fig. S3a. The opposite polarization direction (red and blue arrows) in neighbouring parallel walls cancel out and the macroscopic polarization vanishes.

Flexoelectricity is described by a flexoelectric coefficient $\mu_{12}$:

$$P_3 = \mu_{12} \frac{\partial \varepsilon_{11}}{\partial x_3},$$

where $P$ is the flexoelectric polarization and $\varepsilon_{11}$ is the symmetrized elastic strain tensor.

## S2. Validation of the Machine-Learning deep potential (DP)

After training, the performance of the DP model is validated through the comparison of the energies and forces from DFT calculations in the testing dataset. Considering that the DP model have been proved to be in satisfactory agreement with DFT results in a wide range of bulk phase properties (e.g. antiferrodistortive structure, total energies, atomic forces, elastic constants, and phonon dispersion relation) in our recent works[2,3]. Here, we only use the surface and interface configurations to illustrate the accuracy of the DP model. The forces predicted by the DP model versus DFT calculated energies and forces are plotted in Fig. S3, the mean absolute error (MAE) of energy $|E^{DFT}-E^{DP}|$ and atomic force $|f^{DFT}-f^{DP}|$ between DP and DFT is 0.805 meV/atoms and 0.037 eV/Å, respectively, suggesting the accuracy of the trained DP model. In addition, the DP and DFT calculated (0 0 1) surface formation energies of freestanding structures with different thickness ($N$) are shown in Fig. S8, their differences are very small, at the level of sub-meV/atom. Table SI summarizes equilibrium surface structural properties optimized by DP and DFT at 0 K, the surface structure for the DP prediction agrees well with DFT result, again demonstrating the accuracy of DP model.

We also use the DP model to calculate the various domain wall of $SrTiO_3$, including easy wall, hard wall, and twin wall. For the easy wall, the wall normal to the axis of octahedron rotation (see Fig. S9a). We constructed a supercell of 200 atoms (10 cubic cells in length) elongated along the oxygen octahedron rotation axis ($c$ axis). While for the hard wall, wall parallel to axis of octahedron rotation (see Fig. S9b), and we built a 400-atom supercell (20 cubic cells in length) elongated perpendicular to the oxygen octahedron rotation axis. For the twin wall, it occurs between two tetragonal domains whose respective rotation axis are oriented at 90° with respect to each other (see Fig. 1a). The DP predicted domain wall energies are 5.75 mJ/m$^2$ (twin wall), 24.58 mJ/m$^2$ (hard wall), and 14.63 mJ/m$^2$ (easy wall), which are very close to the DFT values of 6.35 mJ/m², 26.30 mJ/m², and 15.06 mJ/m². In addition, The DP model reproduces the same TiO$_6$ octahedron rotation pattern and $P$ profiles as DFT results (Fig. 1b, c), demonstrating the DP model can describe the various domain walls' structure in DFT accuracy, though the training dataset does not contain any structural information of the domain wall configurations.

## S3. Atomic structure of twin wall

The initial building block of tetragonal phase (optimized structure with a = 3.885 Å and c = 3.906 Å) for supercell with periodic sequence of twin walls is shown in Fig. S1a. Since two TWs are needed to implement periodic boundary conditions, the supercell with periodic sequence of twin walls was constructed by blocks to elongated along a [1 0 1] direction perpendicular to the TW, and one unit cell along directions of *[-1 0 1]* and [1 0 1] parallel to the TWs. Two parallel twin walls are derived using twin-like operations with head-to-head or head-to-tail patterns in supercell (Fig. S2a and b). Finally, the fully relaxed 1×1×20 supercell with two twin walls in a projection along the [0 1 0] axis is shown in Fig. S2a. In supercell, the same color of TiO$_6$ octahedrons indicate that they have the same orientation of rotation. The exaggerated schematic of the supercell about the (0 1 0) planes is shown in Figure 1a. The wall energy of TW is 6.35 mJ/m², which is much lower than 26.30 mJ/m² for antiphase hard wall and 15.06 mJ/m² for easy wall[13], demonstrating that 90° TW is the most stable wall configuration in SrTiO$_3$. The supercell shows a periodic structural corrugation, as illustrated in Fig.1a. The concave (180.58°) and convex (179.42°) twin angles are formed at two antisymmetric TWs about (-1 0 1) plane. These twin angles also cause corrugations on the (0 0 1) surface (see Fig. S3d). Such spontaneous surface corrugation originates from lattice matching of the different extension of their tetragonal unit cells at TWs, as sketched in Fig.2b. Therefore the angle α can be estimated from the geometry with a and c lattices ratio: tan(α)=$c/a$. Interestingly, such (0 0 1) surface corrugation can be revealed by optical reflection microscopy as a striped contrast in experiment[14] and then its angle, tan(α)=0.001, has been indirectly measured by Honig et al[15]. According to calculations, whether the TW is the concave or convex in surface only depends on the orientation of tetragonal elongate axes respect to the walls, and the four possible scenarios with alternate between concave and convex are sketched in Fig. S4.

## S4. Polarization property of twin wall

In order to gain a deeper understanding on the polarity in TWs, we calculate the local unit-cell polarization ($P$) components profiles by atomic displacements and Born effective charges (see Methods), as shown in Fig. 1c. The result shows a slight coupling between two antisymmetric TWs even when a 400-atom supercell was used in DFT calculations. To prevent the coupling between periodic sequence of parallel TWs, a larger supercell is needed, but this is difficult to calculate within the length-scale of DFT. The well-trained DP model allows us to perform the large-scale atomic simulations. More details of DP model training and validation can be found in the SI and Figure S3 and S4, as well as our previous work. The DP model reproduces the same $\varphi$ and $P$ profiles as DFT results (Fig. 1b, c), demonstrating that the DP model can describe the TW's structure with DFT accuracy, even though the training dataset does not contain any structural information of the TW configuration. The DP model was adopted for the structural optimization and the optimized 200-, 400-, and 800-atom supercells with TWs and their corresponding polarization profiles are shown in Fig. S5, which indicates that an 800-atom supercell is sufficiently large to prevent coupling between two TWs. The TWs are divided into two categories based on the transition of axial vector $\varphi$ in wall: head to head (HH) and head to tail (HT) types as sketched by arrows in Fig. S2a, b. Our calculations indicate HH and HT walls have the same properties including TW width, $P$ components magnitude, and $P$ vector direction (Fig. S2c, d). This result is inconsistent with the previous Hamiltonian model results that indicate $|P|_{HH} > |P|_{HT}$ [16]. Therefore, it may be unrealistic to generate a macroscopic ferrielectric-like polarization in a periodic alternating HT and HH walls, as mentioned in previous literature based on the model Hamiltonian.

|  | DFT | DP |
|---|---|---|
| Lattice $a$ (Å) | 7.7906 | 7.7971 |
| Angle of surface Ti-O-Ti (°) | 174.505 | 174.490 |
| Thickness (Å) | 17.4543 | 17.4500 |
| Angle of surface Sr-O-Sr (°) | 171.023 | 171.001 |

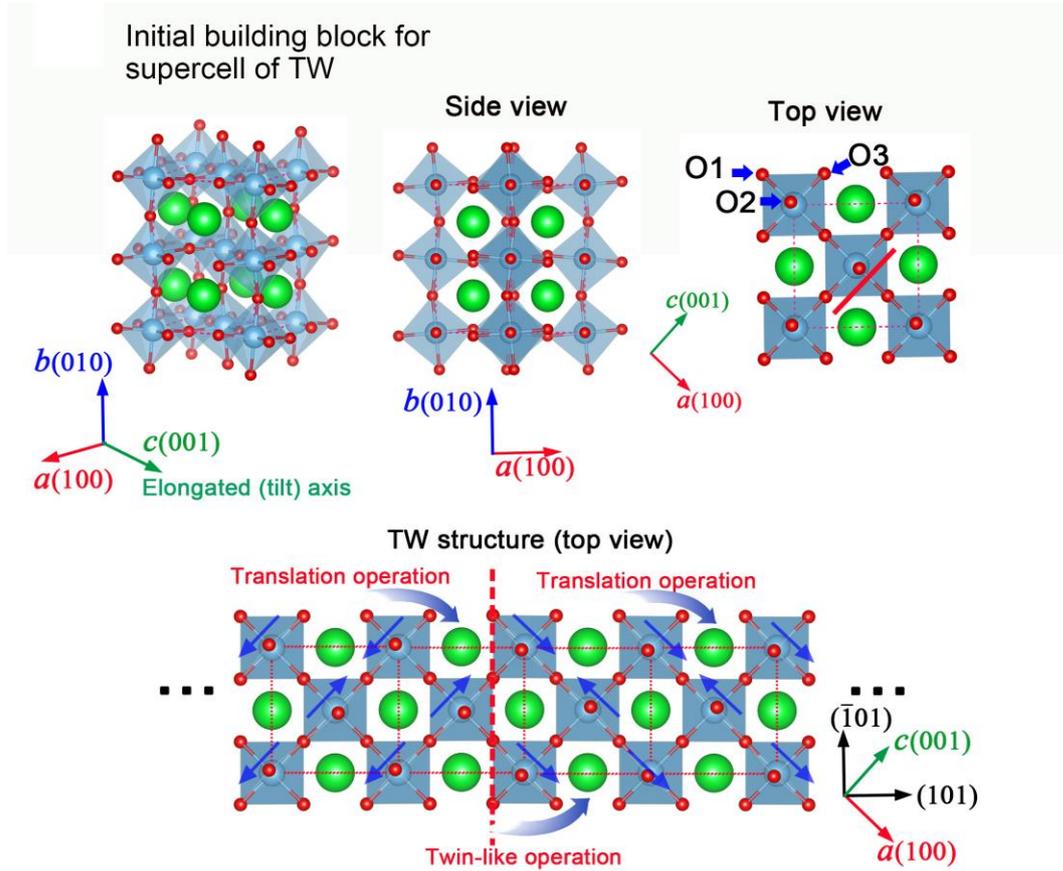

Figure S1 (a) Building block of tetragonal SrTiO$_3$ perovskite for supercells with 90° twin walls construction. (b) Geometry of supercells with twin wall and the TW was created by using a twin-like operation. The blue arrows represent the axial vector $\varphi$ from the staggered rotations of TiO$_6$ octahedra.

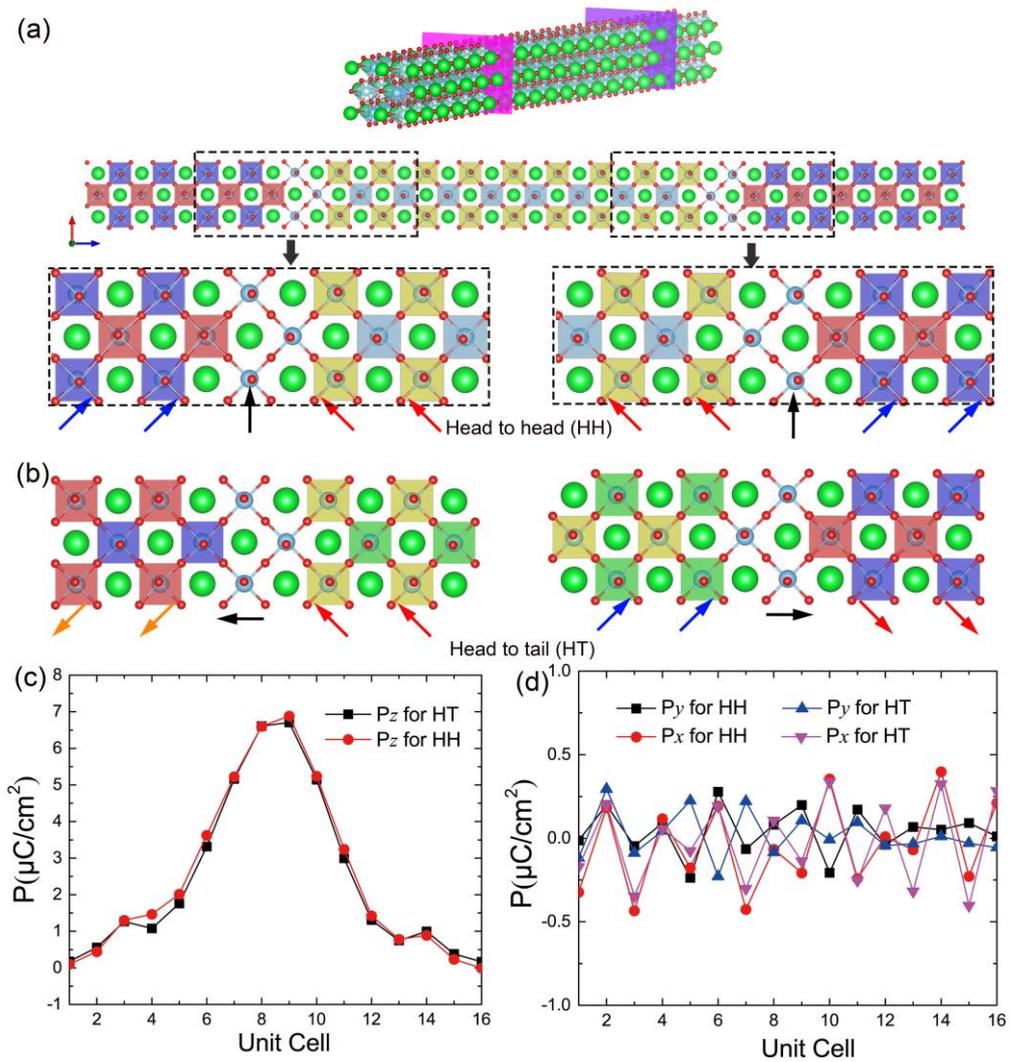

Figure S2 DFT relaxed supercells with (a) head-to-head and (b) head to tail twin walls. (c) Polarization component ($P_x$, $P_y$, and $P_z$) profiles depending on the distance across the two antisymmetric twin walls.

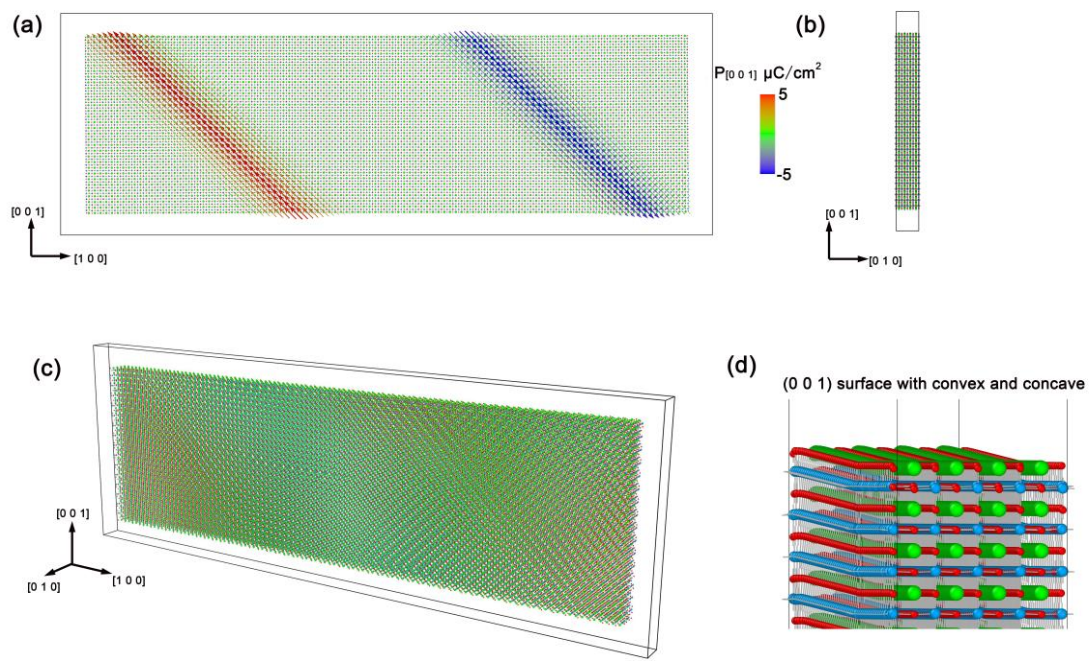

Figure S3 (a-c) Atomistic structure and distribution of dipole configurations in SrTiO3 membrane with 106 × 31 × 4 pseudo cubic units. The two twin walls are parallel to the (1 0 1) plane. (d) The visualization of delicate convex and concave corners in (0 0 1) surface.

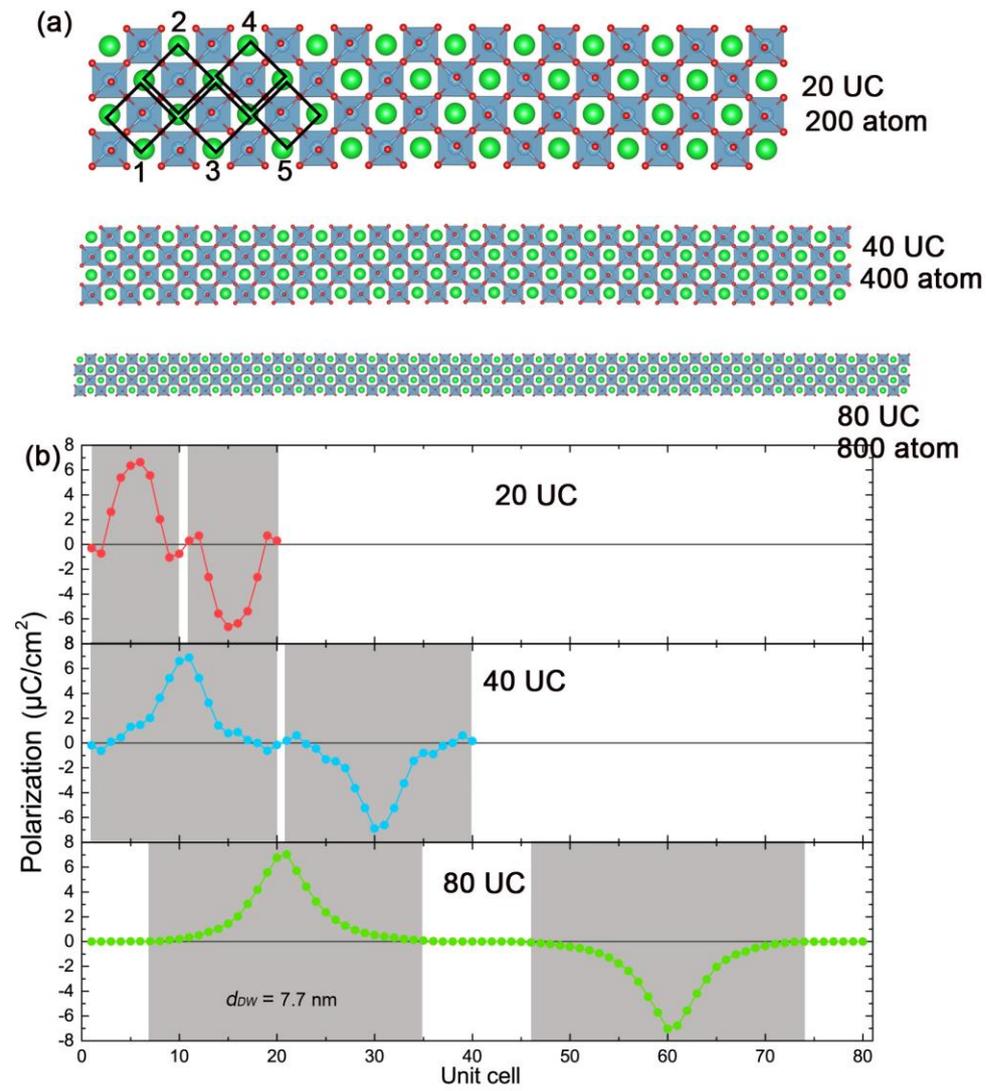

Figure S4 (a) Geometry of $1 \times 1 \times n$ ($n = 10$, 20, and 40) supercells with two twin walls. (b) Polarization profiles in supercells with different length.

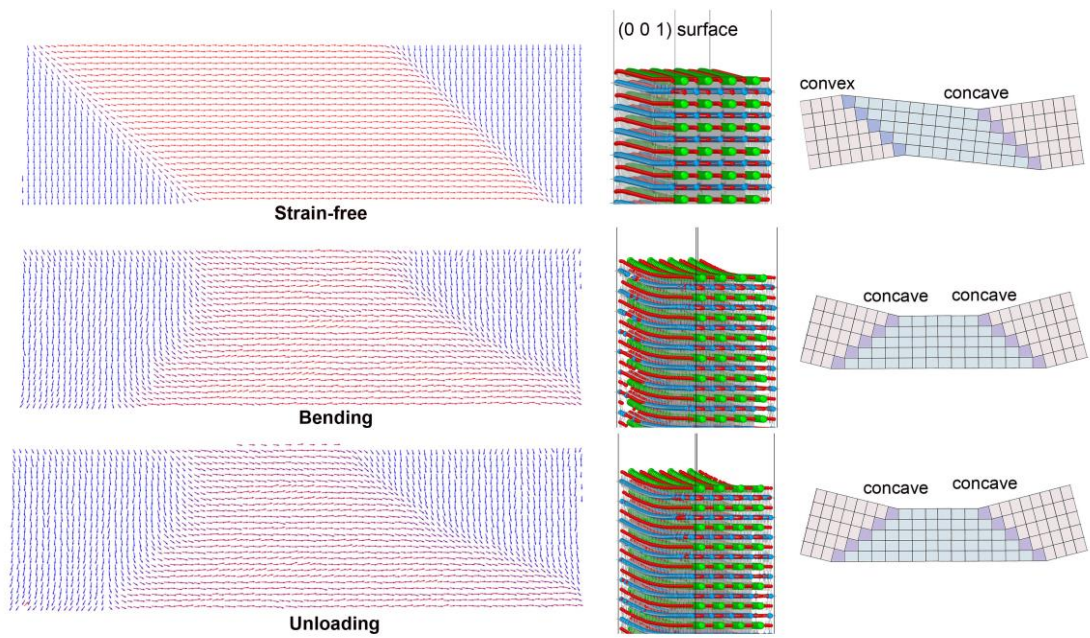

Figure S5 The distribution of axial vector of staggered rotations of TiO6 octahedra (φ) configurations at strain-free, bending, and unloading states.

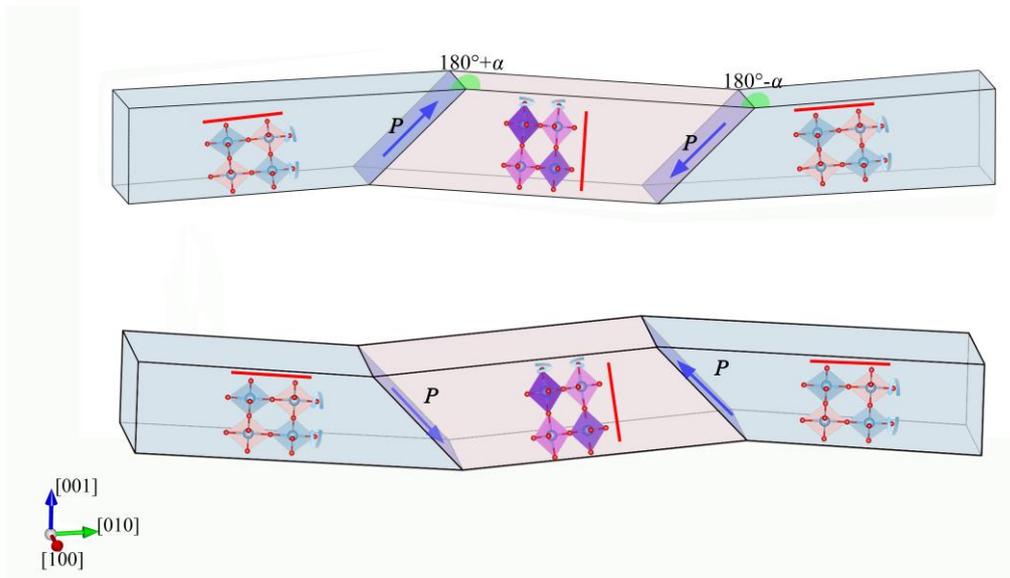

Figure S6 Schematic representation of four possible scenarios with alternate between concave and convex in (0 0 1) surface of SrTiO₃ membrane, indicating the concave or convex in surface depends on the orientation of tetragonal elongate axes respect to the walls.

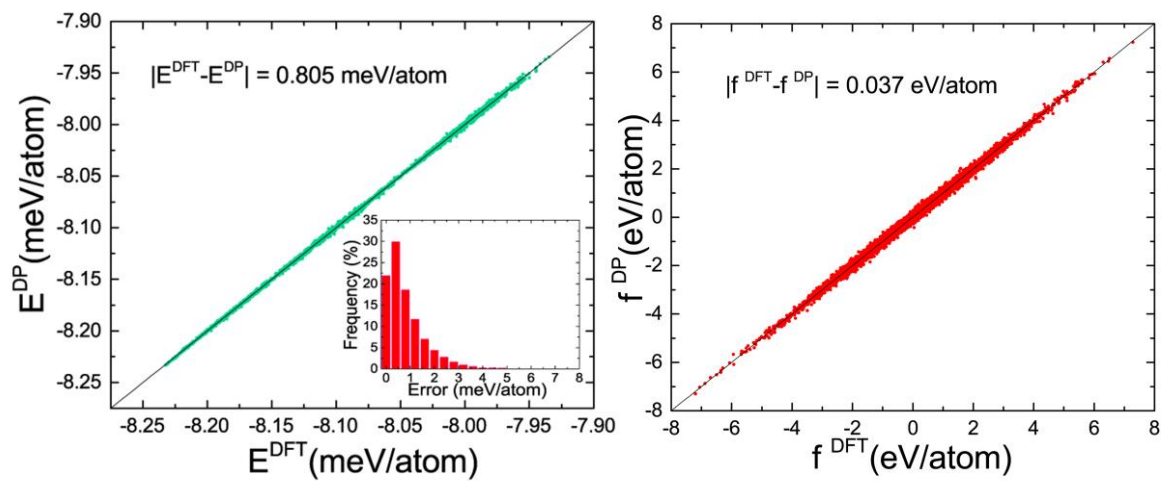

Figure S7 Comparison of energies (a) and atomic force (b) of the DP model against DFT calculations for the freestanding configurations in the training dataset.

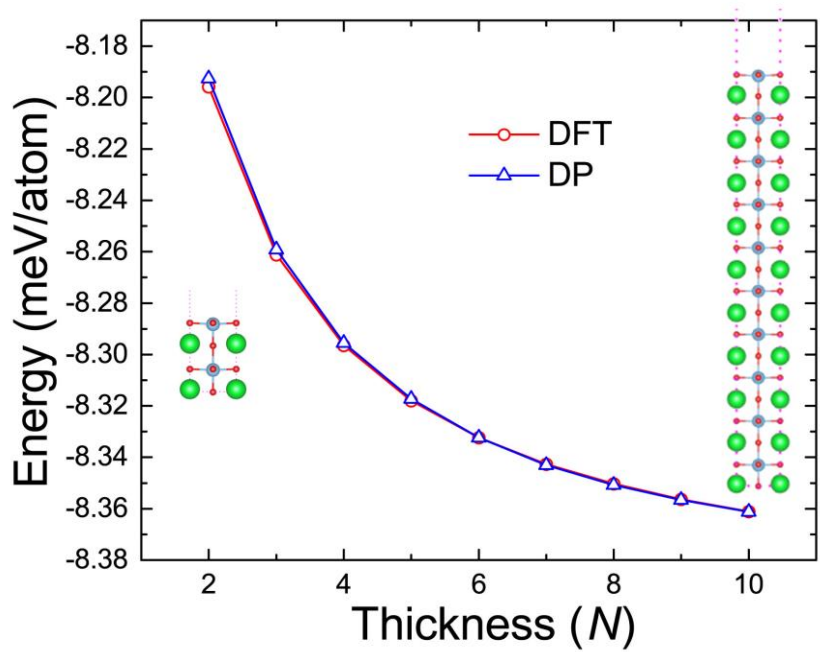

Figure S8 The DP and DFT calculated (0 0 1) surface formation energies of freestanding structures with different thickness (*N*)

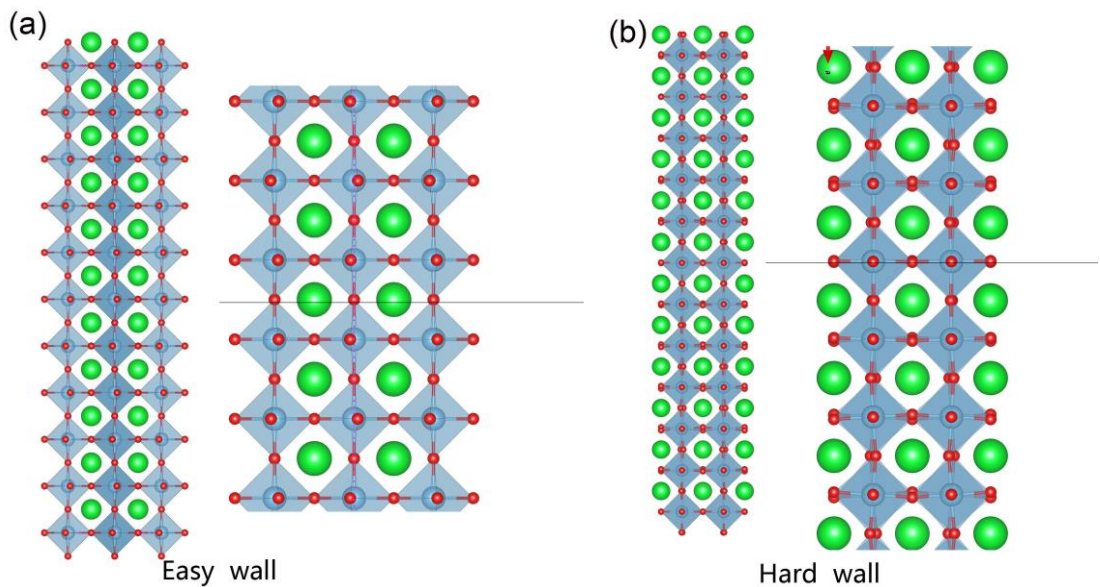

(a) Easy wall

(b) Hard wall

Figure S9 Schematics for the matching of oxygen octahedra between two translational domains in tetragonal SrTiO3 separated with an easy (a) and a hard (b) wall. The rotation of octahedra is about the caxis.

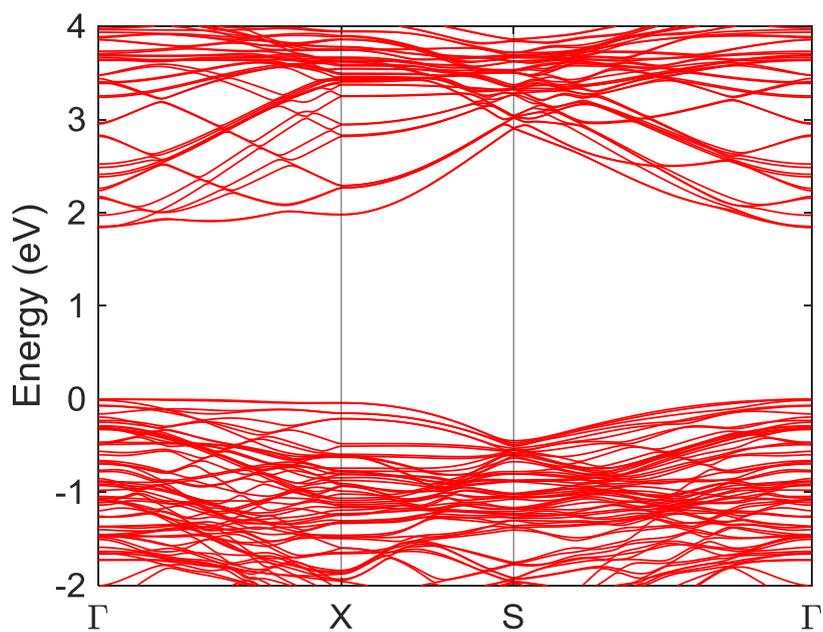

Figure S10. Electronic band structure of the STO supercell used to calculate the BPV conductivities shown in Figure 5 of the main text.